\documentclass[10pt,conference]{IEEEtran}
\IEEEoverridecommandlockouts
\usepackage{cite}
\usepackage{amsmath,amssymb,amsfonts}
\usepackage{algorithmic}
\usepackage{multirow}
\usepackage{subcaption}
\usepackage{graphicx}
\usepackage{textcomp}
\usepackage{xcolor}
\PassOptionsToPackage{hyphens}{url}\usepackage{hyperref}
\usepackage{wrapfig}

\usepackage{fancyvrb}

\RecustomVerbatimCommand{\VerbatimInput}{VerbatimInput}%
{fontsize=\footnotesize,
	frame=lines,  
	framesep=2em, 
	label=\fbox{data.txt},
	labelposition=topline,
	commentchar=*        
}

\newcommand{\codecse}{CodeCSE}
\newcommand{\codesearch}{code search}
\newcommand{\codesearchnet}{CodeSearchNet}

\usepackage{enumitem}
\newlist{questions}{enumerate}{2}
\setlist[questions,1]{label=RQ\arabic*.,ref=RQ\arabic*}
\setlist[questions,2]{label=(\alph*),ref=\thequestionsi(\alph*)}

\newlist{tasks}{enumerate}{2}
\setlist[tasks,1]{label=Task\arabic*.,ref=Task\arabic*}
\setlist[tasks,2]{label=(\alph*),ref=\thequestionsi(\alph*)}

\usepackage{multicol}
\usepackage{multirow}
\usepackage{array}
    \newcolumntype{P}[1]{>{\centering\arraybackslash}p{#1}}
    \newcolumntype{M}[1]{>{\centering\arraybackslash}m{#1}}

\def\BibTeX{{\rm B\kern-.05em{\sc i\kern-.025em b}\kern-.08em
    T\kern-.1667em\lower.7ex\hbox{E}\kern-.125emX}}
\begin{document}

\title{CodeCSE: A Simple Multilingual Model for \\
       Code and Comment Sentence Embeddings 
    \thanks{This material is based upon work supported by the National Science Foundation under Grant No. 2100050.}
}

\author{
    \IEEEauthorblockN{Anthony Varkey}
    \IEEEauthorblockA{
        Ann Arbor, MI, USA \\
    }
    \and
    \IEEEauthorblockN{Siyuan Jiang}
    \IEEEauthorblockA{\textit{Eastern Michigan University} \\
        Ypsilanti, MI, USA \\
        sjiang1@emich.edu
    }
    \and
    \IEEEauthorblockN{Weijing Huang}
    \IEEEauthorblockA{
        Ann Arbor, MI, USA \\
    }
}

\maketitle

\begin{abstract}
Pretrained language models for code token embeddings are used in code search, code clone detection, and other code-related tasks. Similarly, code function embeddings are useful in such tasks. However, there is no out-of-box models for \textit{function embeddings} in the current literature. So, this paper proposes {\codecse}, a contrastive learning model that learns embeddings for functions and their descriptions in one space.

We evaluated {\codecse} using {\codesearch}. {\codecse}'s multilingual zero-shot approach is as efficient as the models finetuned from GraphCodeBERT for specific languages.

{\codecse} is open source at \href{https://github.com/emu-se/codecse}{https://github.com/emu-se/codecse} and the pretrained model is available at the HuggingFace public hub: \href{https://huggingface.co/sjiang1/codecse}{https://huggingface.co/sjiang1/codecse}
\end{abstract}


\section{Introduction}
Sentence embeddings, which are embeddings of sentences (instead of tokens), are preferred in semantic search tasks in NLP. For example, SentenceBERT and SimCSE~\cite{reimers-2019-sentence-bert, gao-2021-simcse} are two models that provide sentence embeddings for search on general NLP documents. Compared to token embeddings, sentence embeddings can be used out-of-box without further tuning. However, there are no such public models that generate sentence embeddings for code and comments in software engineering. To address this issue, we propose a pretrained model for sentence-level code and comment embeddings using contrastive learning, named \textbf{CodeCSE} (\textbf{Code} and comment \textbf{C}ontrastive \textbf{S}entence \textbf{E}mbedding). 

In recent years, pre-trained models have shown great performance in various code-related sequence generation tasks like code-to-comment and comment-to-code, where the best examples are OpenAI's codex, and ChatGPT. \codecse is different from these models because these pretrained models' outputs are token sequences and \codecse's outputs are embeddings. The embeddings can be used in vector databases, for example, milvus~\footnote{https://milvus.io/}, for fast semantic search.

CodeCSE uses contrastive learning to ensure that code and comment pairs are mapped closer in the embedding space if they match each other (in other words, the comment is the code's comment.) To use contrastive learning, an encoder is needed to generate embeddings. In \codecse, we use GraphCodeBERT~\cite{guo-2020-graphcodebert}, which is a Transformer encoder, as the encoder of the contrastive learning. Only one encoder is used in \codecse, which means the code and comment go through the same encoder, where their embeddings are mapped into the same space. The way of using “contrastive” in CodeCSE can also be seen in other multi-modal representation learning works, e.g., CLIP~\cite{radford-2021-CLIP}. Our contribution is mainly on the experiment side and the ecosystem side. On the experiment side, we run (1) base model selection, CodeBERT or GraphCodeBERT, (2) hyperparameter tunings, such as the pooling method, the number of MLP layers on top of the base model, the loss function, and the other experiment settings, such as the batch size, and the mini-batch construction. And on the ecosystem side, we enriched the code-related NLP tools by contributing an out-of-box pretrained model to generate sentence-level embeddings for code and comments. 

To verify the effectiveness of code and comment embeddings based on CodeCSE, we conducted the experiment focusing on zero-shot code semantic search, where we use the code embeddings to search for semantically related comments and vice versa. The experiment result shows CodeCSE’s cross-language zero-shot approach is as efficient as the models finetuned from GraphCodeBERT for specific languages. 


Coincidentally, OpenAI also used contrastive pretraining for code embeddings. It is a side-product of their main topic, cpt-text~\cite{neelakantan-2022-cpt}. OpenAI trained cpt-code on its in-house datasets and reported the code embeddings could significantly improve the code search result on {\codesearchnet}~\cite{husain2019codesearchnet}. However, the model is not published for public access to this day. This limits further studies and the public use of sentence-level code embeddings. Our {\codecse} model is open-sourced, and the pretrained model is on HuggingFace's public hub\footnote{\href{https://huggingface.co/sjiang1/codecse}{https://huggingface.co/sjiang1/codecse}}. 

\section{Related Work}
\subsection{Pretrained models for code tokens}
Pretrained models have emerged as a powerful tool for improving natural language processing (NLP) tasks. BERT, in particular, has shown impressive performance on a range of downstream NLP tasks, utilizing a pretraining-and-then-finetuning approach. Building on this success, the software engineering community has developed their own pretrained models to handle downstream tasks specific to code. The examples include CuBERT~\cite{kanade2020cubert}, CodeBERT~\cite{feng-2020-codebert}, GraphCodeBERT~\cite{guo-2020-graphcodebert}, and UniXcoder~\cite{guo-2022-unixcoder}, and etc. By focusing on token-level representations, these models are demonstrated effective in code-related downstream tasks, such as code search, code clone detection, code translation~\cite{guo-2020-graphcodebert}, code summarization, and code completion~\cite{guo-2022-unixcoder}.

CodeBERT~\cite{feng-2020-codebert} and GraphCodeBERT~\cite{guo-2020-graphcodebert} are two notable examples of pretrained models for code tokens. In the inference phase, both models use a RoBERTa~\cite{liu-2019-roberta} architecture, but they differ in their data preprocessing. For the code part, GraphCodeBERT generates a data flow representation, which is then appended to the code in its preprocessing, while CodeBERT does not incorporate any additional data into the code. CodeBERT and GraphCodeBERT are trained on multilingual datasets, while CuBERT~\cite{kanade2020cubert} is only trained on Python code.

In terms of pretraining objectives, CodeBERT utilizes two objectives: Masked Language Modeling (MLM) and Replaced Token Detection (RTD). MLM involves randomly masking certain tokens in the input sequence and training the model to predict their original values. RTD replaces some tokens with similar tokens and trains the model to identify which tokens were replaced. CodeBERT is pretrained on NL-PL pairs from six programming languages on the CodeSearchNet dataset, excluding the development and test sets. GraphCodeBERT builds upon CodeBERT's approach by incorporating the data flow information into the pretraining process. In addition to MLM, GraphCodeBERT uses two dataflow-related reconstruction tasks to further improve the model's ability to understand code. These tasks involve predicting the code or dataflow given the other and predicting the masked dataflow given the code.


\subsection{Contrastive learning for code-related task}

Contrastive learning has emerged as a powerful technique for learning representations of multi-modal data by mapping similar objects closer together and dissimilar objects further apart. 
A typical work is CLIP~\cite{radford-2021-CLIP}, mapping a pair of caption text and an image into the same space to be close. 

Comment and code pairs can also be viewed as multi-modal data, and it is natural to apply contrastive learning techniques to learn representations for these pairs. Several models, including SynCoBERT~\cite{wang2021syncobert}, C3P~\cite{C3P}, and CodeRetriever~\cite{li-2022-coderetriever}, have been developed to learn such representations for code and comments
Furthermore, these models are in a pretraining-and-fine-tuning style, so they require fine-tuning for specific downstream tasks, such as code retrieval and summarization. 
The contrastive learning is a part of the loss function in the pre-training process of SynCoBERT, C3P, and CodeRetriever. Meanwhile, the purpose of using contrastive learning in these works is not to provide sentence embeddings for code and comment. 

\cite{wang-2022-multiview-contrastive}
\cite{wang2022heloc}
\cite{li-2022-soft-contrastive}

Despite these limitations, contrastive learning has the potential to significantly improve the quality of representations for code and comments, by leveraging the large amounts of unlabeled data available in software repositories. The ability to learn high-quality embeddings for code and comments would enable a wide range of applications, from code retrieval to topic analysis. Moreover, by using a contrastive learning framework, it may be possible to learn embeddings that are more robust to variations in programming languages and coding styles, further expanding the potential use cases for such models.

\subsection{BERT-based sentence embeddings}
Sentence-BERT~\cite{reimers-2019-sentence-bert} is introduced in the scenario that needs embeddings on a large scale for the semantic search or similarity comparison, such as clustering.  
Before that, BERT takes in the (query, doc) pairs for the ad-hoc computation, which limits its application on a large scale. 
The literature~\cite{Chang2020Pre-training-largescale-retrieval} points out that the embedding-based model can benefit the retrieval stage while BERT-style (query, doc) ad-hoc computation can be used for the scoring stage. 

Based on BERT, Sentence-BERT is continuously trained on a supervised sentence pair dataset with contrastive learning. A widely-used pre-trained Sentence-BERT model is all-mpnet-base-v2\footnote{\href{https://huggingface.co/sentence-transformers/all-mpnet-base-v2}{https://huggingface.co/sentence-transformers/all-mpnet-base-v2}}, which is trained on 1 billion pairs. 
SimCSE~\cite{gao-2021-simcse}, on the other hand, is a self-supervised learning method that learns sentence embeddings without any labeled data. Its contrastive loss aims to maximize the similarity between different encoded versions of the same sentence, while minimizing the similarity between different sentences. 

\subsection{GPT-based sentence embeddings}
OpenAI used contrastive learning for text and code embeddings~\cite{neelakantan-2022-cpt}, namely cpt-text and cpt-code. However the main topic in the paper is cpt-text. They are both based on GPT models and use the last token's embedding in the last layer as the sentence embedding. 
The cpt-code is initialized with codex~\cite{chen-2021-codex}. 
The early codex versions are trained on 54 million public GitHub software repositories and fine-tuned on pre-trained GPT-3~\cite{brown2020gpt3} model family. 
Though cpt-code~\cite{neelakantan-2022-cpt} showed that its training target is the contrastive loss on (text, code) pairs extracted from open source code, it didn't give the details about its in-house dataset's size and programming languages covered in the dataset used to train cpt-code. 

SGPT~\cite{muennighoff2022sgpt} used the open-sourced GPT-Neo~\cite{gpt-neox-library}, and GPT-J~\cite{gpt-j} as the backbone to provide the sentence embedding for text.  
It uses position-weighted mean pooling on the last layer rather than the last token. 
It also adapts the BitFit~\cite{ben-zaken-2022-bitfit} method to only tune the GPT models' bias parameters to make the finetuning more memory efficient. 

Both cpt-text and SGPT report that they outperform BERT-based sentence embeddings in text. 
Since there's no public pre-trained GPT models on code, we leave the open-sourced GPT-based embeddings on code as future work.

\subsection{Reproducibility of related works}
\label{sec:relatedworks:reproducibility}
The ability to reproduce research results enables other researchers to verify the findings, build upon them, and compare different models or methods. 
We listed the reproducibilities of the related works in Table~\ref{tab:relatedworks_reproducibility}. 
This table covered six aspects. 
\begin{enumerate}
    \item model structure. It refers to the source code defining the architecture of the model, including the embedding layer, transformer layers, and any potential top layers. 
    \item model weights. Model weights are binary files (e.g., pytorch\_model.bin) that contain the learned parameters of the model, which are used to initialize the model during inference or fine-tuning. The model weights need to match the model structure.
    \item pretraining script. It refers to the source code that defines the entire pretraining process, including the source of the pretraining dataset, dataset preprocessing, self-supervised pretraining tasks, loss functions, optimization method, the training parameters, and etc. It enables other researchers to pretrain the model on different datasets or retrain the model from scratch. 
    \item finetuning script. It refers to the source code that defines the fine-tuning process, including how to fine-tune the pre-trained model for specific downstream tasks, such as code search or code summarization. Usually, it includes the data source of specific tasks, data preprocessing, training parameters, and etc. 
    \item evaluation script. It's the source code that defines the process of evaluating the performance of the models on specific downstream tasks. It includes how to load the fine-tuned model for specific tasks, how to preprocess the test data, and how to calculate the evaluation metrics. 
    \item inference (deployment) script. The inference (deployment) script defines the process of using the fine-tuned model to generate predictions on new data. It includes how to tokenize the input (e.g., a vocabulary file), how to parse the data into a format that the model can read (e.g., how to parse the "code" part into a data flow and append it to the input), and how to generate the final output for the inference (deployment). 
\end{enumerate}
We considered the availability of the above aspects in the authors' paper and on repositories such as GitHub and HuggingFace Hub.

Of the related works we evaluated, CodeBERT and GraphCodeBERT stand out for their high level of reproducibility. Both models released their model structures, model weights, and scripts for finetuning, evaluation, and deployment. CodeBERT\footnote{\href{https://huggingface.co/microsoft/codebert-base}{https://huggingface.co/microsoft/codebert-base}}'s model structure and weights have been downloaded about 1.6 million times in February 2023, and GraphCodeBERT has served as the foundation for further research in academia~\cite{C3P, li-2022-coderetriever, wang-2022-multiview-contrastive}. For example, CodeRetriever~\cite{li-2022-coderetriever} is initialized with GraphCodeBERT's checkpoint.

Ensuring the reproducibility of our proposed model, CodeCSE, is a priority for us. 
We followed the same six aspects listed in Table~\ref{tab:relatedworks_reproducibility} to make CodeCSE fully reproducible. 
Specifically, we will release code and data, provide detailed instructions for pretraining and finetuning, and document evaluation results. By doing so, we hope to contribute to the advancement of research and practical applications in the field of machine learning for code and comments.

\begin{table*}[htbp]
	\begin{center}
\begin{tabular}{|l|cccccc|}
\hline
               & \multicolumn{1}{c|}{\begin{tabular}[c]{@{}c@{}}model \\ structure\end{tabular}} & \multicolumn{1}{c|}{\begin{tabular}[c]{@{}c@{}}model \\ weights\end{tabular}} & \multicolumn{1}{c|}{\begin{tabular}[c]{@{}c@{}}pretraining \\ script\end{tabular}} & \multicolumn{1}{c|}{\begin{tabular}[c]{@{}c@{}}fine-tuning \\ script\end{tabular}} & \multicolumn{1}{c|}{\begin{tabular}[c]{@{}c@{}}evaluation \\ script\end{tabular}} & \begin{tabular}[c]{@{}c@{}}inference (deploy) \\ script\end{tabular} \\ \hline
CodeBERT~\cite{feng-2020-codebert}       & \multicolumn{1}{c|}{\checkmark}                                  & \multicolumn{1}{c|}{\checkmark}                                & \multicolumn{1}{c|}{no}                                                            & \multicolumn{1}{c|}{\checkmark}                                     & \multicolumn{1}{c|}{\checkmark}                                    & \checkmark                                            \\ \hline
GraphCodeBERT~\cite{guo-2020-graphcodebert}  & \multicolumn{1}{c|}{\checkmark}                                  & \multicolumn{1}{c|}{\checkmark}                                & \multicolumn{1}{c|}{no}                                                            & \multicolumn{1}{c|}{\checkmark}                                     & \multicolumn{1}{c|}{\checkmark}                                    & \checkmark                                            \\ \hline
SynCoBERT~\cite{wang2021syncobert}      & \multicolumn{6}{c|}{no repo}                                                                                                                                                                                                                                                                                                                                                                                                                                                                         \\ \hline
C3P~\cite{C3P}            & \multicolumn{1}{c|}{\checkmark}                                  & \multicolumn{1}{c|}{\checkmark}                                & \multicolumn{1}{c|}{no}                                                            & \multicolumn{1}{c|}{no}                                                            & \multicolumn{1}{c|}{no}                                                           & no                                                                   \\ \hline
CodeRetriever~\cite{li-2022-coderetriever}  & \multicolumn{6}{c|}{empty repo @ 97fd33c}                                                                                                                                                                                                                                                                                                                                                                                                                                                            \\ \hline
cpt-code~\cite{neelakantan-2022-cpt} & \multicolumn{6}{c|}{no repo}                                                                                                                                                                                                                                                                                                                                                                                                                                                            \\ \hline 
CodeCSE (Our work) & \multicolumn{1}{c|}{\checkmark}                                  & \multicolumn{1}{c|}{\checkmark}                                & \multicolumn{1}{c|}{\checkmark}                                     & \multicolumn{1}{c|}{\checkmark}                                     & \multicolumn{1}{c|}{\checkmark}                                    & \checkmark                                            \\ \hline
\end{tabular}
	\caption{The reproducibilities of related works and CodeCSE}
	\label{tab:relatedworks_reproducibility}
	\end{center}
\end{table*}

\section{Background and Preliminaries}
\subsection{Contrastive Learning}
Contrastive learning is a way to generate lower dimensional representations of inputs. This learning method has been applied in various domains with different input types. For example, it is used to generate reprensentations of images~\cite{radford-2021-CLIP} and sentences~\cite{gao-2021-simcse}. The learned representations are embeddings---lower dimensional vectors.

\begin{figure}[htbp]
    \centering
    \includegraphics[width=0.4\linewidth]{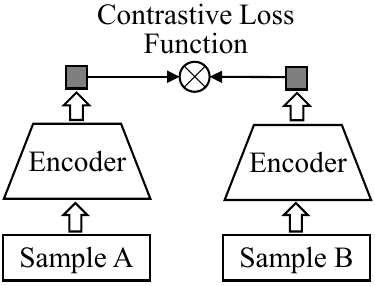}
    \caption{Overview of contrastive learning.}
    \label{fig:contrastive_learning}
\end{figure}

Figure~\ref{fig:contrastive_learning} illustrates an overview of contrastive learning. The input of a contrastive learning model is a pair of samples, e.g., Sample A and Sample B in Figure~\ref{fig:contrastive_learning}. A binary label is assigned to the pair of samples based on whether they are neighbors, which is provided by the training dataset. Then, the two samples are encoded independently into two embeddings. In some models, the two encoders share parameters, which means the two smaples are encoded by one encoder. In other models, the two encoders are independent. After getting the embeddings of the two samples, the model uses a contrastive loss function to calculate the training loss based on the assigned label (whether the samples are neighbors). This loss function pulls neighbor samples' embeddings close to each other.~\cite{Hadsell-2006-constrastive,gao-2021-simcse}. In this way, the neighborhood relationships among the samples are preserved in the learned embeddings.

\subsubsection{Unimodal, Bimodal, and Multimodal}
Sample A and Sample B in Figure~\ref{fig:contrastive_learning} can be in the same data type or in different data types. For example, to learn embeddings of images and their captions, Sample A can be an image and Sample B can be a text~\cite{radford-2021-CLIP}. Unimodal, bimodal, and multimodal are three scenarios of combining sample types.

\begin{itemize}
    \item Unimodal: Sample A and Sample B are two samples of the same data type, e.g., code. Unimodal positive samples in code do not naturally exist in code databases. To find similar code functions, for example, CodeRetriever uses a two-step approach: 1) collecting function pairs through function name match and document match, and 2) training and using a classifer model to filter noisy function pairs~\cite{li-2022-coderetriever}.
    \item Bimodal: Sample A and Sample B are two samples of different types, e.g., code and natural language. Bimodal positive samples in code naturally exist in code databases, which are functions and their documents. This is the common setup in learning code embeddings with contrastive learning.
    \item Multimodal: The samples can be more than two types. The input for the contrastive learning is still a pair of samples. But one of the samples can be more than one type. For example, SynCoBERT's sample types include 1) NL, 2) code with AST, 3) NL with code with AST (in two different orders)~\cite{wang2021syncobert}.
\end{itemize}

\subsubsection{Strategies for Selecting Negative Samples}

The quality of negative samples affects the performance of contrastive learning~\cite{kanade2020cubert,gao-2021-simcse,li-2022-coderetriever}. Here, we list common negative sampling strategies.

\begin{itemize}
    \item Using \textbf{in-batch negative} samples. In practice, a batch of training examples is a set of positive pairs. For a batch of $n$ paired samples, $\{(A_{i}, B_{i})\: \vert\: i = [1..n] \}$, for each $A_{i}$, the positive sample is $B_{i}$. The in-batch negative samples are $\{B_{j}\}\: \text{where}\: j \neq i$.
    \item Using \textbf{hard negative} samples. Hard negatives are the representative samples that may be more difficult to be learned. For example, a contradiction of a sentence is a hard negative compared to a random unrelated sentence. In some domains, manually labeled hard negative samples are available,which can improve the performance of contrastive learning~\cite{gao-2021-simcse}.
\end{itemize}

\subsection{Transformers, BERT, RoBERTa, and Pre-training}
Our model, {\codecse}, and all the baselines in this paper use transformer encoders. A transformer encoder is a series of layers; each layer contains two sublayers: a multi-head self-attention layer and a position-wise feed-forward layer~\cite{vaswani2017attention}. The input and output of each layer has the same size. If a Transformer encoder has 768 dimensional hidden states and a maximum sequence length of 512, this encoder's input should be a sequence of 512 tokens. Each token has an embedding of 768 dimensions. The output of this encoder is also a sequence of 512 tokens with embeddings of 768 dimensions.

Additionally, {\codecse} and all the baselines follow the pretraining framework in BERT~\cite{devlin-2018-bert}, where the input sequence is split into two segments using three special tokens: $[CLS]$, $[SEP]$, and $[EOS]$. $[CLS]$ denotes the beginning of the sequence; $[SEP]$ is the delimitor of the two segments; and $[EOS]$ denotes the end of the sequence. Given an input sequence, $[[CLS], x_1, ..., x_N, [SEP], y_1, ..., y_M, [EOS]]$, the output embedding for $[CLS]$ is traditionally considered as a summary embedding for segment $[x_1, ..., x_N]$ and segment $[y_1, ..., y_M]$. If only one segment is given, the second segment is set to empty and $[CLS]$ is the summary embedding for segment $x_1, ..., x_N$.

\subsection{Sentence Embedding}
In contrast to token embeddings which are representations of tokens, a sentence embedding is a representation of a sentence. This embedding is supposed to capture the semantic information of a sentence as a whole. In this paper, we use the term ``sentence embedding'' to describe an embedding of a piece of code or an embedding of a natural language document. In our pretraining and evaluation, the piece of code is a code function and its corresponding NL document is the function's comment.
\section{CodeCSE}
This section describes the model architecture and pretraining process of {\codecse}. Overall, {\codecse} is a simple bi-modal contrastive learning model using a transformer as its encoder for both code and NL documents. Figure~\ref{fig:codecse_model} illustrates {\codecse}'s model architecture. 

\subsection{Encoder}
\begin{figure}[htbp]
    \centering
    \includegraphics[width=0.7\linewidth]{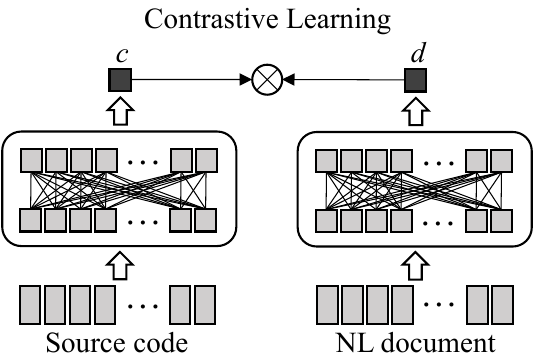}
    \caption{{\codecse}'s dual-encoder model architecture. Source code and NL documents are encoded independently by an encoder. The encoder for source code and the encoder for NL documents share parameters. The encoder outputs are fed into contrastive learning. In contrastive learning, the outputs are pooled into $c$ and $d$, which are the sentence embeddings for source code and the NL document respectively.}
    \label{fig:codecse_model}
\end{figure}
During training, the inputs of the model are $\langle \text{source code}$, $\text{NL document} \rangle$ pairs. No special treatement is done for document inputs. For code, {\codecse} uses GraphCodeBERT's method to integrate data flow graph information in inputs using an attention mask~\cite{guo-2020-graphcodebert}. The mask is for the encoder to avoid performing attention on irrelevant tokens.

{\codecse} has one encoder, which encodes code and documents independently. This encoder is a 12-layer transformer with 768 dimensional hidden states and 12 attention heads. For contrastive learning, each encoder output is pooled into one sentence embedding of 768 dimensions ($c$ and $d$ in Figure~\ref{fig:codecse_model}).

\subsection{Pooling}
\begin{figure}[htbp]
    \centering
    \includegraphics[width=0.4\linewidth]{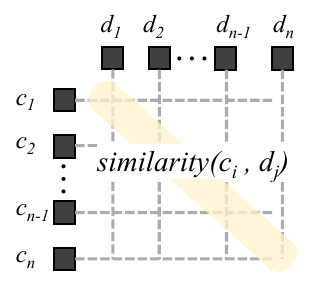}
    \caption{{\codecse}'s contrastive learning after pooling. The contrastive learning module takes a minibatch of pairs of code and document embeddings as inputs. $n$ is the size of the minibatch. ($c_i$, $d_j$) is a pair of sentence embeddings of a piece of code and a document. {\codecse} calculates the similarity of ($c_i$, $d_j$) for any $i$ and $j$ in $[1,...,n]$. If $i$ equals to $j$, the pair is a positive sample. Otherwise, the pair is a negative sample. The training loss is calculated based on the similarities.}
    \label{fig:codecse_cl}
\end{figure}
First, {\codecse} pools the encoder's outputs into sentence embeddings ($c$ and $d$ in Figure~\ref{fig:codecse_model}). {\codecse} provides four options for pooling: 
\begin{enumerate}
    \item {\textit{cls}}: using the [CLS] representation as the sentence embedding.
    \item {\textit{avg}}: using the average of all the tokens' hidden states at the last layer as the sentence embedding.
    \item {\textit{avg\_first\_last}}: using the average of all the tokens' hidden states at the first and the last layers as the sentence embeddings.
\end{enumerate}

\subsection{MLP layers after pooling}
After pooling, we get a sentence embedding for each input (code or document). To add more trainable parameters for sentence embeddings, {\codecse} has options to put one or more MLP layers after pooling. Each MLP layer is a linear layer with a Tangent activation function.

\subsection{Training loss}
After pooling and the optional MLP layers, each minibatch is a set of code sentence embeddings and their documents' sentence embeddings. Then, {\codecse} calculates the cosine similarities between code and document embeddings, illustrated in Figure~\ref{fig:codecse_cl}. {\codecse} uses the in-batch negative sampling. Each code and its document forms a positive sample. Each code and any document that is not the code's document forms a negative sample. For a minibatch size of $n$, for each code sentence embedding, $c_i$, there is one positive sample ($c_i$, $d_i$) and $n-1$ negative samples ($c_i$, $d_j$) where $j \neq i$.

{\codecse} uses cross entropy on the cosine similarities to calculates the loss for all the $c_i$ (Equations~\ref{eq:loss:cidj} and~\ref{eq:loss:ci}) and the loss for all the $d_j$ (Equations~\ref{eq:loss:djci} and~\ref{eq:loss:dj}).
\begin{equation}
    \label{eq:loss:cidj}
    \text{loss}(c_i, d_j) = -\log \frac{\exp(similarity(c_i, d_j))}{\Sigma_{k=1}^{n} \exp(similarity(c_i, d_k))}
\end{equation}
\begin{equation}
    \label{eq:loss:ci}
    \text{loss}(c_i) = \Sigma_{j=1}^{n} \frac{\text{loss}(c_i, d_j)}{n}
\end{equation}

\begin{equation}
    \label{eq:loss:djci}
    \text{loss}(d_j, c_i) = -\log \frac{\exp(similarity(c_i, d_j))}{\Sigma_{k=1}^{n} \exp(similarity(c_k, d_j))}
\end{equation}
\begin{equation}
    \label{eq:loss:dj}
    \text{loss}(d_j) = \Sigma_{i=1}^{n} \frac{\text{loss}(d_j, c_i)}{n}
\end{equation}

Finally, {\codecse} has three options to calculate the training loss based on $\text{loss}(c_i)$ and $\text{loss}(d_j)$:

\begin{enumerate}
    \item {\textit{symmetric}}: using the average of the loss for code and the loss for documents, Equation~\ref{eq:loss:sim}.
    \begin{equation}
        \label{eq:loss:sim}
        \text{symmetric\_loss}(c_i, d_j) = \frac{\text{loss}(c_i) + \text{loss}(d_j)}{2}
    \end{equation}
    \item {\textit{asymmetric\_code}}: 
        using $\text{loss}(c_i)$.
    \item {\textit{asymmetric\_doc}}: 
        using $\text{loss}(d_j)$. 
\end{enumerate}

\subsection{Pretraining}
To accelerate the training process, {\codecse} is initialized with GraphCodeBERT-base's parameters\footnote{\href{https://huggingface.co/microsoft/graphcodebert-base}{https://huggingface.co/microsoft/graphcodebert-base}}. Then, we sampled each batch from the same programming language because the random negative samples from other programming languages are too easy for the model to learn. An oversampling strategy~\cite{conneau-2019-cross-lingual,guo-2020-graphcodebert} is applied in pretraining so that the low-resource languages are sampled more often. The probability of a programming language being sampled is adjusted according to Equation~\ref{eq:adjusted_probability}, where $q_i$ is the ajudsted probability of the $i$-th programming language and $p_i$ is the fraction of the programming language's samples in the whole dataset.

\begin{equation}
    \label{eq:adjusted_probability}
    q_i = {\frac{p_{i}^{\alpha}}{\Sigma_{k=1}^{N} {p_{k}^{\alpha}}} \; \text{where} \; \alpha = 0.7}
\end{equation}

To have a simpler implementation, in each epoch, the batches of one programming languages are not mixed up with other programming languages. For example, the order of the programming languages of the batches in the first epoch is Java, Ruby, Python, Go, JavaScript, and PHP. The languages are rotated after each epoch so that the last language becomes the first in the next epoch. So, in the second epoch, the order of the languages becomes PHP, Java, Ruby, Python, Go, and JavaScript. The batches are re-sampled in the beginning of every epoch, so that each batch has different negative samples across different epochs.

In ablation studies, we trained {\codecse} for only one epoch for each setting to save energy cost. In evaluation, we trained {\codecse} for 12 epochs to obtain the state-of-the-art results.
\section{Ablation Studies}
\label{sec:ablation}

In the ablation studies, we examined the effects of the following design choices: 1) pooler types, 2) MLP layers, and 3) loss functions on contrastive learning. We kept the batch size to 96 for all the models in the ablation studies. 96 is the largest batch size we can use on an Nvidia 40GB A100 for {\codecse}. We trained the models for one epoch to reduce the energy cost of training models.

\subsection{Research questions}

\begin{questions}[itemindent=1em]
\item What is the impact of pooler types on the generated embeddings? \label{rq:pooler}
\item What is the impact of the number of MLP layers on the generated embeddings? \label{rq:mlp}
\item What is the impact of loss functions on the generated embeddings? \label{rq:loss}
\end{questions}

\subsection{Datasets}
\begin{table}[htbp]
	\begin{center}
	\begin{tabular}{lrr}
	\hline
			       & \multicolumn{1}{c}{Train}   & \multicolumn{1}{c}{Valid}        \\ \hline
	Ruby       & 48,791  & 2,209      \\ 
	JavaScript & 123,889  & 8,253     \\ 
	Java       & 454,451  & 15,328    \\ 
	Go         & 317,832  & 14,242    \\ 
	PHP        & 523,712  & 26,015    \\  
	Python     & 412,178  & 23,107    \\  \hline
	Total      & 1,880,853 & 89,154  \\ \hline
	\end{tabular}
	\caption{Bimodal datasets in CodeSearchNet. }
	\label{tab:dataset:ablation}
	\end{center}
\end{table}

We used the bimodal part of the training and validation sets in the CodeSearchNet~\cite{husain2019codesearchnet} for training and evaluation. To minimize the risk of overfitting, we did not use the test sets for the ablation studies. By doing so, we can accurately gauge the impact of our design choices on the model's performance without introducing any bias from the test dataset. The dataset statistics are shown in Table~\ref{tab:dataset:ablation}. We obtained the datasets from Hugging Face datasets.\footnote{\href{https://huggingface.co/datasets/code\_search\_net}{https://huggingface.co/datasets/code\_search\_net} Accessed in August 2022.}

\subsection{Metrics}
We used the alignment score to describe the quality of the generated embeddings. The alignment score was proposed by Wang and Isola~\cite{wang-2022-alignment} for contrastive learning evaluation, and was adopted for eavluating other contrastive learning models~\cite{gao-2021-simcse}. This score describes the closeness of paired embeddings in Euclidean space.

Equation~\ref{eqn:alignment} defines the alignment score. The function $f(\cdot)$ produces normalized embeddings of code or text, such that $\|f(\cdot)\|=1$. Here, $n$ is the total number of pairs. $f({x}_i)$ is the embedding of a code function, and $f({x}^+_i)$ represents the embedding of a document.

\begin{equation}
  \label{eqn:alignment}
  l_{\text{align}} = \frac{1}{n}\sum_{i=1}^{n}\| f({x}_i) - f({x}^+_i)\|^2
\end{equation}

If all the ${x}_i$ and ${x}^+_i$ are positive pairs (the document is the code function's summary), the alignment score should be lower, indicating the positive pairs have embeddings closer to each other in Euclidean space. In contrast, the alignment score for negative pairs should be higher. The difference between these two alignment scores represents the closeness of positive pairs compared to negative pairs. 

For each programming language, we calculated an alignment score for all the positive pairs, and an alignment score for 50,000 random pairs from the validation set. In the rest of the paper, we denote the alignment score for positive pairs as $\text{align}\textsubscript{positive}$, the alignment score for random pairs as $\text{align}\textsubscript{negative}$, and the difference between the $\text{align}\textsubscript{positive}$ and the $\text{align}\textsubscript{negative}$ as $\text{align}\textsubscript{diff}$.

\subsection{Results}
\begin{table*}[htbp]
  \begin{center}
    \begin{tabular}{P{1.5cm}P{1cm}P{2.5cm}P{2.5cm}P{2.5cm}}
      \hline
      \multirow{2}{1.5cm}{\centering pooler \\ type} & \multirow{2}{1cm}{\centering MLP \\ layer} & \multicolumn{3}{c}{loss function}\\
      \cline{3-5}
       & & $asymmetric\_code$ & $asymmetric\_doc$ & $symmetric$ \\
      \hline
      \multirow{3}{1.5cm}{\centering cls} & 0 & 0.716, 1.938, 1.222 & 0.743, 1.949, 1.206 & 0.759, 1.948, 1.189  \\ 
       & 1 & 0.711, 1.949, 1.238  & 0.737, 1.957, 1.219 & 0.754, 1.955, 1.201       \\
       & 2 & 0.698, 1.953, 1.255  & 0.722, 1.952, 1.230 & 0.742, 1.953, 1.211       \\
      \hline
      \multirow{3}{1.5cm}{\centering avg} & 0 & 0.775, 1.951, 1.176 & 0.817, 1.974, 1.157 & 0.851, 1.985, 1.134 \\ 
       & 1 & 0.760, 1.957, 1.197 & 0.794, 1.971, 1.177 & 0.830, 1.983, 1.153 \\
       & 2 & 0.734, 1.953, 1.219 & 0.769, 1.963, 1.194 & 0.790, 1.967, 1.177 \\
      \hline
      \multirow{3}{1.5cm}{\centering $avg$\\$\_first\_last$} & 0 & 0.782, 1.950, 1.169 & 0.823, 1.975, 1.151 & 0.855, 1.987, 1.132 \\
       & 1 & 0.761, 1.954, 1.193 & 0.792, 1.969, 1.178 & 0.827, 1.981, 1.154 \\
       & 2 & 0.735, 1.951, 1.216 & 0.763, 1.960, 1.197 & 0.787, 1.965, 1.177 \\
      \hline
    \end{tabular}
    \caption{Ablation studies results of all the model configurations. Each cell has three scores: $\text{align}\textsubscript{positive}$, $\text{align}\textsubscript{negative}$, and $\text{align}\textsubscript{diff}$.  $\text{Align}\textsubscript{positive}$ is the alignment score of positive pairs. $\text{Align}\textsubscript{negative}$ is the alignment score of random pairs. $\text{Align}\textsubscript{diff}$ is the difference between $\text{align}\textsubscript{positive}$ and $\text{align}\textsubscript{negative}$. All the models were pretrained for one epoch with a batch size of 96.}
    \label{tab:ablation}
  \end{center}
\end{table*}

\begin{figure*}[htbp]
  \centering
  \begin{subfigure}{0.33\textwidth}
    \centering
    \includegraphics[width=.95\linewidth]{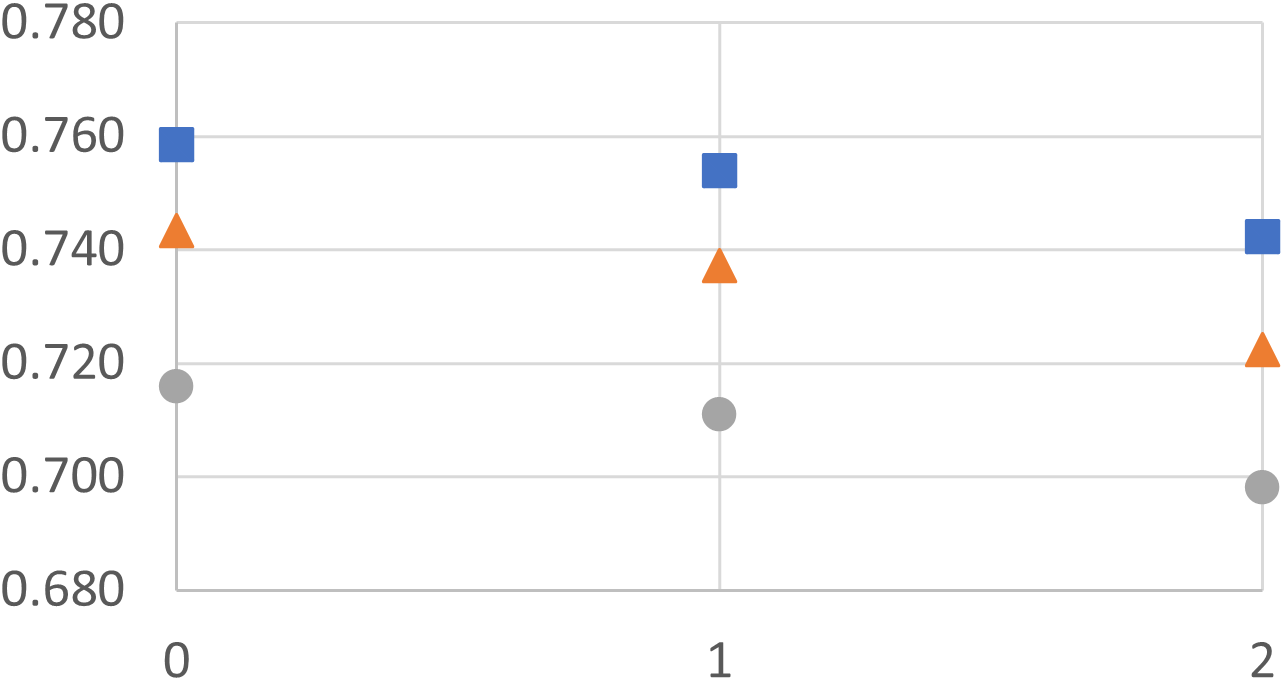}
    \caption{$\text{Align}\textsubscript{positive}$ scores for models with $cls$.}
    \label{fig:ablation_alignpos}
  \end{subfigure}
  \begin{subfigure}{0.32\textwidth}
    \centering
    \includegraphics[width=.95\linewidth]{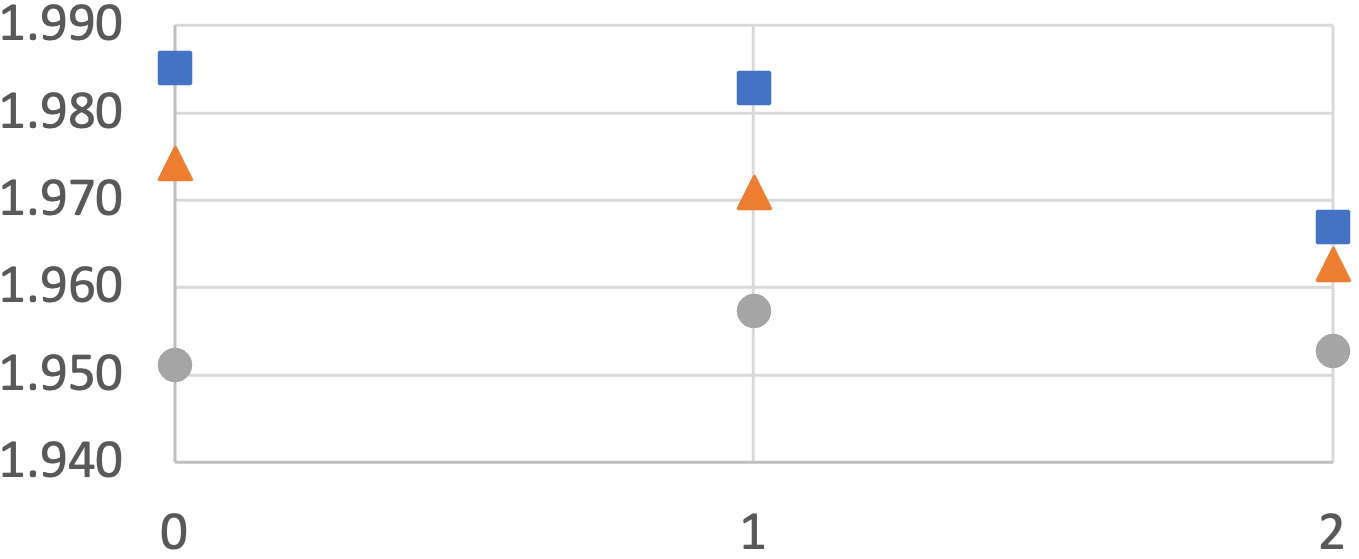}
    \caption{$\text{Align}\textsubscript{negative}$ scores for models with $avg$.}
    \label{fig:ablation_aligndrand_avg}
  \end{subfigure}
  \begin{subfigure}{0.33\textwidth}
    \centering
    \includegraphics[width=.95\linewidth]{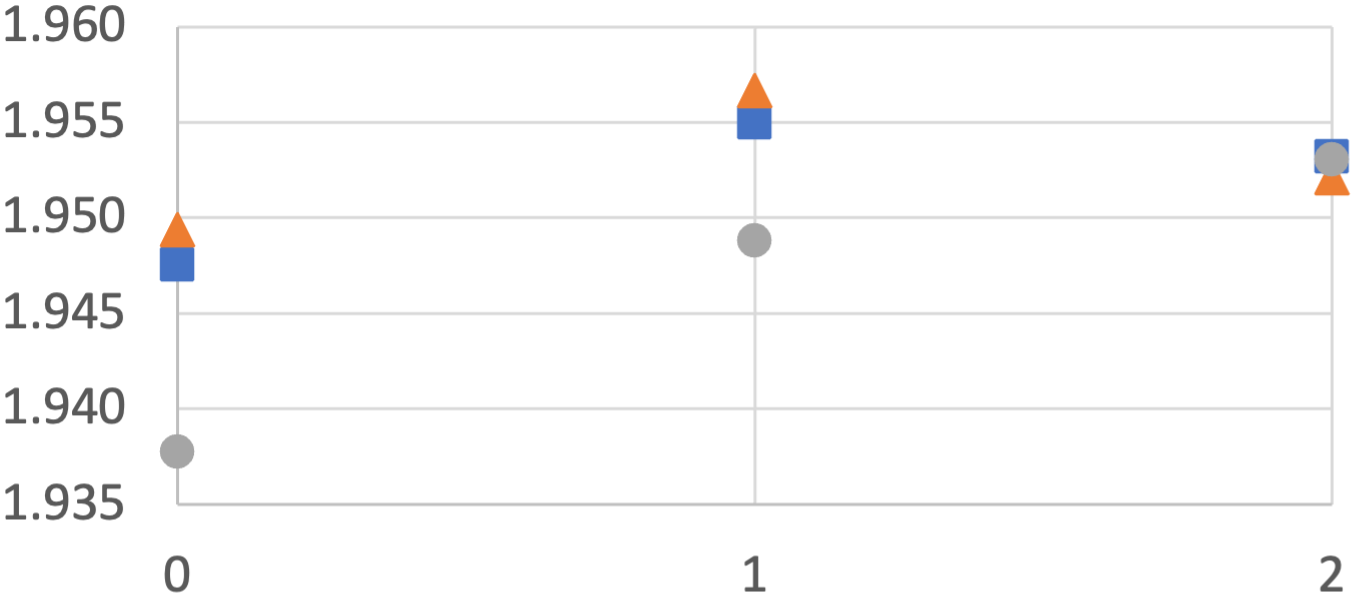}
    \caption{$\text{Align}\textsubscript{negative}$ scores for models with $cls$.}
    \label{fig:ablation_alignrand}
  \end{subfigure}

  \caption{Alignment scores in the ablation studies. Rectangles denote $symmetric$ loss, triangulars denote $asymmetric\_doc$, and circles denote $asymmetric\_code$.}
  \label{fig:ablation}
\end{figure*}

Table~\ref{tab:ablation} shows the results of all the configurations: three pooler types ($cls$, $avg$, $avg\_first\_last$), three numbers of MLP layers (0, 1, and 2), and three loss functions ($symmetric$, $asymmetric\_code$, and $asymmetric\_doc$). In total, we trained 27 models with all the possible configurations.

For each model, for each metric, we report the mean value of all the programming langauges. For the results of all the models, the mean of $\text{align}\textsubscript{positive}$ is 0.771 with a standard deviation of 0.115. The mean of $\text{align}\textsubscript{negative}$ is 1.961 with a standard deviation of 0.031. This indicates that the distances of positive pairs' embeddings vary among different models more than the distances of random pairs' embeddings.

\subsubsection{\ref{rq:pooler} The impact of the pooler type}
The models with $cls$ have lower $\text{align}\textsubscript{positive}$, and higher $\text{align}\textsubscript{diff}$. This indicates that, compared to other poolers, $cls$ brings the embeddings of positive pairs closer.

\subsubsection{\ref{rq:mlp} The impact of the number of MLP layers}
The models with more MLP layers have lower $\text{align}\textsubscript{positive}$, and higher $\text{align}\textsubscript{diff}$. Figure~\ref{fig:ablation_alignpos} shows the results of the models with the $cls$ pooler. With a $cls$ pooler, for all the loss functions, models with two MLP layers have lower $\text{align}\textsubscript{positive}$ scores and higher $\text{align}\textsubscript{diff}$ scores. The models with other poolers have the same trends about the number of MLP layers. 

On the other hand, more MLP layers lead to shorter distances among random pairs.  Figure~\ref{fig:ablation_aligndrand_avg} shows the results of the models with the $avg$ pooler. The models with $symmetric$ loss and $asymmetric\_doc$ loss have lower $\text{align}\textsubscript{negative}$ if they have more MLP layers. For $asymmetric\_code$, one MLP layer has the highest $\text{align}\textsubscript{negative}$ score, and two MLP layers has a slightly lower $\text{align}\textsubscript{negative}$ score.



\subsubsection{\ref{rq:loss} The impact of the loss function}
The loss function of $asymmetric\_code$ results in lower $\text{align}\textsubscript{positive}$ scores, but also has lower $\text{align}\textsubscript{negative}$ scores. Figure~\ref{fig:ablation_alignrand} shows the $\text{align}\textsubscript{negative}$ scores of the models with the $cls$ pooler. 

Based on $\text{align}\textsubscript{diff}$, $asymmetric\_code$ is the best performing loss function, and $symmetric$ is the worst loss function.

\graphicspath{{./images}}
\section{Using {\codecse} on Code Search}

\subsection{Task and datasets}
To see how well the embeddings generated by {\codecse} can work on a code search task, we used the task defined in GraphCodeBERT~\cite{guo-2020-graphcodebert} and used their code search dataset prepared from CodeSearchNet~\cite{husain2019codesearchnet}. GraphCodeBERT cleaned the dataset for the code search task, so this dataset is different from the original CodeSearchNet dataset. The statistics of this dataset is listed in Table~\ref{tab:dataset:code_search}. The input of the task is a natural language query, and the search candidates are codes (functions). The goal is to find out the functions that are most semantically similar to the query.

The evaluation metric is MRR (Mean Reciprocal Rank). MRR reflects the rank of the ground truth in a ranking list. MRR scores closer to 1 mean that the ground truth is ranked higher on average across all queries.

\begin{table*}[htbp]
	\begin{center}
	\begin{tabular}{lrrrrrrrr}
	\hline
			   & \multicolumn{2}{c}{Train}              & \multicolumn{2}{c}{Valid}             & \multicolumn{2}{c}{Test}              & \multicolumn{2}{c}{Candidates} \\  \hline
	Ruby       & \multicolumn{1}{r}{24,927}  & 2.75\%   & \multicolumn{1}{r}{1,400}  & 3.13\%   & \multicolumn{1}{r}{1,261}  & 2.40\%   & \multicolumn{1}{r}{4,360}  & 2.38\% \\ 
	JavaScript & \multicolumn{1}{r}{58,025}  & 6.39\%   & \multicolumn{1}{r}{3,885}  & 8.69\%   & \multicolumn{1}{r}{3,291}  & 6.26\%   & \multicolumn{1}{r}{13,981} & 7.63\% \\ 
	Java       & \multicolumn{1}{r}{164,923} & 18.17\%  & \multicolumn{1}{r}{5,183}  & 11.60\%  & \multicolumn{1}{r}{10,955} & 20.84\%  & \multicolumn{1}{r}{40,347} & 22.01\% \\ 
	Go         & \multicolumn{1}{r}{167,288} & 18.43\%  & \multicolumn{1}{r}{7,325}  & 16.39\%  & \multicolumn{1}{r}{8,122}  & 15.45\%  & \multicolumn{1}{r}{28,120} & 15.34\% \\ 
	PHP        & \multicolumn{1}{r}{241,241} & 26.58\%  & \multicolumn{1}{r}{12,982} & 29.05\%  & \multicolumn{1}{r}{14,014} & 26.66\%  & \multicolumn{1}{r}{52,660} & 28.73\% \\ 
	Python     & \multicolumn{1}{r}{251,820} & 27.68\%  & \multicolumn{1}{r}{13,914} & 31.14\%  & \multicolumn{1}{r}{14,918} & 28.38\%  & \multicolumn{1}{r}{43,827} & 23.91\% \\ \hline
	Total      & \multicolumn{1}{r}{907,624} & 100.00\% & \multicolumn{1}{r}{44,689} & 100.00\% & \multicolumn{1}{r}{52,561} & 100.00\% & \multicolumn{1}{r}{183,295}& 100.00\% \\ \hline
	\end{tabular}
	\caption{Dataset of Code Search}
	\label{tab:dataset:code_search}
	\end{center}
\end{table*}

\begin{figure}
	\centering
	\includegraphics[width=\linewidth]{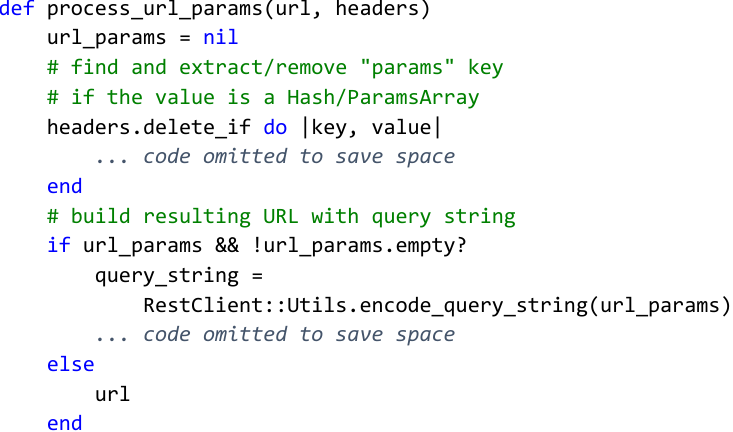}
	\caption{A Ruby function in the validation set for the code search task. Its corresponding query is ``Extract the query parameters and append them to the url''. The comments in the function are removed in the preprocessing step.}
	\label{fig:codesearch:example}
\end{figure}

For example, an NL query in the validation set for Ruby is ``Extract the query parameters and append them to the url''. The ground truth is the function in Figure~\ref{fig:codesearch:example}. CodeCSE ranked this function the first place among all the 4,360 Ruby functions. The Reciprocal Rank in this case is $1$.

\subsection{Pretrained models}
\label{sec:codesearch:models}
In this section, to get state-of-the-art performance, we pretrained {\codecse} for 12 epochs, with a batch size of 960, using eight Nvidia 80GB A100 GPUs. Two {\codecse} models were pretrained:

\begin{enumerate}
	\item $\text{avg\_mlp0\_symmetric}$: the model has an avg pooler, zero MLP layer, and a $symmetric$ loss function. 
	\item $\text{cls\_mlp2\_asymmetric}$: the model has a cls pooler, two MLP layers, and an $asymmetric_{code}$ function. 
\end{enumerate}
	
In the ablation study, the second configuration works significantly better than the first configuration. We report the alignment scores for these two models in Table~\ref{tab:fullytrained_alignment}. Under the same configuration, both models achieved better alignment scores because they had more epochs in training and a larger batch size.

\begin{table}[htbp]
	\begin{center}
		\begin{tabular}{P{3cm}P{1.3cm}P{1.5cm}P{1.2cm}}
			\hline
			Model & 
			$\text{Align}\textsubscript{positive}$ & 
			$\text{Align}\textsubscript{negative}$ & 
			$\text{Align}\textsubscript{diff}$ \\ \hline
			$\text{avg\_mlp0\_symmetric}$ & 0.809 & 1.990 & 1.180 \\ 
			$\text{cls\_mlp2\_asymmetric}$ & 0.683 & 1.969 & 1.287 \\
			\hline
        \end{tabular}
	\caption{Alignment scores for $\text{avg\_mlp0\_symmetric}$ and $\text{cls\_mlp2\_asymmetric}$. Both models were pretrained for 12 epochs with 960 batch size. Alignment scores are defined in Equation~\ref{eqn:alignment}, Section~\ref{sec:ablation}.}
	\label{tab:fullytrained_alignment}
	\end{center}
\end{table}

\subsection{Approach}
\textbf{We used the zero-shot learning to apply CodeCSE on the {\codesearch} task}, meaning that we did not fine-tune the model on the (query, code) pairs. We used CodeCSE directly to generate embeddings for query and code, and calculated cosine similarities of code and query embeddings to get a similarity score. All the inputs were preprocessed in the same way we did in pretraining. 

Additionally, \textbf{CodeCSE is a multilingual model} that can generate embeddings for six programming languages, so only one {\codecse} model was used for all programming languages. In contrast, all the baselines' results were obtained by language-specific models.

\subsection{Baselines}
In this experiment, we selected state-of-the-art language models that are designed and pretrained specifically for code. The baselines can be categorized into two groups: 1) models without contrastive learning: CodeBERT~\cite{feng-2020-codebert} and GraphCodeBERT~\cite{guo-2020-graphcodebert}, 2) models with contrastive learning: SynCoBERT~\cite{wang2021syncobert}, C3P~\cite{C3P}, and CodeRetriever~\cite{li-2022-coderetriever}. \textbf{All the baselines, including the models with contrastive learning, are finetuned based on the (query, code) pairs for each specific programming language.}

The encoders of all the baselines and {\codecse} are 12-layer Transformer encoders with 768 dimensional hidden states and 12 attention heads. Table~\ref{tab:eval:codesearch:baselines} lists the use of contrastive learning, the use of unimodal data, model initialization, the number of pretraining steps, batch sizes, and the finetuning approaches of the baselines and CodeCSE. More details about the baselines for the {\codesearch} task are listed below.
\begin{enumerate}
    \item CodeBERT: In CodeBERT's paper, the authors reported that they finetuned a multi-lingual model for six programming languages but that model performed worse than finetuned language-specific models. Originally, CodeBERT used a binary classification approach for finetuning, but later, CodeBERT used the same finetuning process as GraphCodeBERT~\cite{guo-2020-graphcodebert}. 
    \item GraphCodeBERT: Inner products of NL queries' embeddings and codes' embeddings were used to calculate loss in finetuning and inference. Because the samples in a batch serve as negative samples in finetuning, batch generation and the batch size affect the final result. GraphCodeBERT does not have special treatment for batch generation, uses ``In-Batch Negative'' sampling and the batch size is 32. 
    \item SynCoBERT: Uni-modal and Multi-modal contrastive learning was used in pretraining. Paired samples for contrastive learning have three types: NL vs PL-AST, NL-PL-AST vs NL-AST-PL, and PL-AST vs AST-PL~\cite{wang2021syncobert}. The finetuning method is not specified in the paper.
    \item C3P: C3P is the only model with \textbf{two independent encoders (without sharing parameters)}: one encoder for code and the other encoder for NL documents. Both encoders are 12 layers Transformers with 768 dimensional hidden states and 12 attention heads~\cite{C3P}. Bimodal contrastive learning was used in pretraining. The training samples are NL-PL pairs. 
    \item CodeRetriever: Unimodal and bimodal contrastive learning was used in pretraining~\cite{li-2022-coderetriever}. Paired samples for contrastive learning have three types: (1) Document vs PL, (2) In-line comment vs PL, and (3) PL vs PL. Different from other baselines, CodeRetriever has three finetuning approaches: In-Batch Negative, Hard Negative, and AR2. AR2 is an approach to select hard samples using adversarial training~\cite{zhang-2022-ar2}. 
\end{enumerate}

\begin{table*}[htbp]
\begin{center}
\setlength{\tabcolsep}{0.4em}
\begin{tabular}{llp{1.4cm}p{3cm}p{2cm}p{4.2cm}}
\hline
& \begin{tabular}[c]{@{}l@{}}Contrastive learning\\ during pretraining\end{tabular} & Use of unimodal data$^*$ & Model initialization & Pretraining steps $\times$ batch size & Finetuning appraoch \\
\hline
CodeBERT      & No & Yes & RoBERTa's parameters  & 100k $\times$ 2048 & Same as GraphCodeBERT \\
GraphCodeBERT & No & No  & CodeBERT's parameters & 200k $\times$ 1024 & Cross entropy loss over a softmax layer of the inner products of NL and PL embeddings in a batch.    \\ 
SynCoBERT     & Multi-modal, in-batch & Yes & CodeBERT's parameters & 110k $\times$ 128 & Not specified                                                                                             \\
              & negative samples & & & & \\
C3P           & Bi-modal, in-batch & No & CodeBERT's parameters & 512k $\times$ 1024 & Two fully connected layers and the \\ 
              & negative samples & & & & last layer outputs a normalized probability over all search candidates \\
CodeRetriever & Uni-, Multi-modal, & Yes & GraphCodeBERT's & 100k $\times$ 256 & Three sampling approaches. Extra \\ 
& in-batch negative samples & & parameters & & layers are not reported/specified. \\ \hline
CodeCSE       & Bi-modal, in-batch & No & GraphCodeBERT's & 26k $\times$ 960 & No finetuning \\
              & negative samples   &    & parameters      &                  & \\ \hline
\end{tabular}
\caption{The baselines in the {\codesearch} task}
\label{tab:eval:codesearch:baselines}
\end{center}
\end{table*}

\subsection{Results}
\begin{table*}[htbp]
	\begin{center}
	  \setlength{\tabcolsep}{0.4em}
	  \begin{tabular}{P{3.5cm}P{1.7cm}P{1.7cm}P{1.7cm}P{1.7cm}P{1.7cm}P{1.7cm}P{1.7cm}}
		\hline
			& Ruby                    & JavaScript              & Go                      & Python                  & Java                    & PHP                     & Overall\\
		\hline
		$\text{avg\_mlp0\_symmetric}$
		    & 0.770                   & 0.695                   & 0.908                   & 0.721                   & 0.727                   & 0.672                   & 0.749 \\
		$\text{cls\_mlp2\_asymmetric}$
		    & 0.767 (-0.003)          & 0.689 (-0.006)          & 0.908 (0.000)           & 0.721 (0.000)           & 0.722 (-0.005)          & 0.662 (-0.010)                  & 0.745 (-0.004) \\
		\hline 
	  \end{tabular}
	  \caption{The MRR scores of the two {\codecse} models.}
	  \label{tab:eval:codesearch-codecse}
	\end{center}
\end{table*}

\begin{table*}[htbp]
	\begin{center}
	  \setlength{\tabcolsep}{0.4em}
	  \begin{tabular}{P{3.5cm}P{1.7cm}P{1.7cm}P{1.7cm}P{1.7cm}P{1.7cm}P{1.7cm}P{1.7cm}}
		\hline
			& Ruby                    & JavaScript              & Go                      & Python                  & Java                    & PHP                     & Overall\\
		\hline
		CodeBERT$^{\mathrm{*}}$~\cite{feng-2020-codebert}
		    & 0.679 (\textbf{-0.091}) & 0.620 (\textbf{-0.075}) & 0.882 (\textbf{-0.026}) & 0.672 (\textbf{-0.049}) & 0.676 (\textbf{-0.051}) & 0.628 (\textbf{-0.044}) & 0.693 (\textbf{-0.056})\\
		GraphCodeBERT$^{\mathrm{*}}$~\cite{guo-2020-graphcodebert}
			& 0.703 (\textbf{-0.067}) & 0.644 (\textbf{-0.051}) & 0.897 (\textbf{-0.011}) & 0.692 (\textbf{-0.029}) & 0.691 (\textbf{-0.036}) & 0.649 (\textbf{-0.023}) & 0.713 (\textbf{-0.036})\\
		\hline
		SynCoBERT$^{\mathrm{*}}$~\cite{wang2021syncobert}
		    & 0.722 (\textbf{-0.048}) & 0.677 (\textbf{-0.018}) & 0.913 (0.005)           & 0.724 (0.003)           & 0.723 (\textbf{-0.004}) & 0.678 (0.006)           & 0.740 (\textbf{-0.009})\\
		C3P$^{\mathrm{*}}$~\cite{C3P}
		    & 0.756 (\textbf{-0.014}) & 0.677 (\textbf{-0.018}) & 0.906 (\textbf{-0.002}) & 0.704 (\textbf{-0.017}) & 0.759 (0.032)           & 0.684 (0.012)           & 0.748 (\textbf{-0.001})\\
		{\codecse}$^{\mathrm{***}}$
		    & 0.770                   & 0.695                   & 0.908                   & 0.721                   & 0.727                   & 0.672                   & 0.749 \\
		CodeRetriever$^{\mathrm{*}}$ (In-Batch$^{\mathrm{**}}$)~\cite{li-2022-coderetriever} 
			& 0.753 (\textbf{-0.017}) & 0.695 (0.000)           & 0.916 (0.008)           & 0.733 (0.012)           & 0.740 (0.013)           & 0.682 (0.010)           & 0.753 (0.004) \\
		CodeRetriever$^{\mathrm{*}}$ (Hard$^{\mathrm{**}}$)~\cite{li-2022-coderetriever} 
			& 0.751 (\textbf{-0.019}) & 0.698 (0.003)           & 0.923 (0.015)           & 0.740 (0.019)           & 0.749 (0.022)           & 0.691 (0.019)           & 0.759 (0.010) \\
		CodeRetriever$^{\mathrm{*}}$ (AR2$^{\mathrm{**}}$)~\cite{li-2022-coderetriever} 
			& 0.771 (0.001)           & 0.719 (0.024)           & 0.924 (0.016)           & 0.758 (0.037)           & 0.765 (0.038)           & 0.708 (0.036)           & 0.774 (0.025) \\
		\hline
		\multicolumn{8}{p{17cm}}{$^{\mathrm{*}}$The baseline results reported are from previous papers~\cite{guo-2020-graphcodebert, wang2021syncobert, C3P, li-2022-coderetriever}. Since we used GraphCodeBERT's setup, the CodeBERT's results are from GraphCodeBERT's paper~\cite{guo-2020-graphcodebert}.} \\ 
		\multicolumn{8}{p{17cm}}{$^{\mathrm{**}}$ In-Batch Negative, Hard Negative, and AR2 are three finetuning approaches used by CodeRetriever.} \\
		\multicolumn{8}{p{17cm}}{$^{\mathrm{***}}$ The {\codecse} model here is $\text{avg\_mlp0\_symmetric}$.}
	  \end{tabular}
	  \caption{The MRR scores of the baselines and {\codecse} on {\codesearch}.}
	  \label{tab:eval:codesearch}
	\end{center}
\end{table*}

\subsubsection{Comparing the two {\codecse} models}
Table~\ref{tab:eval:codesearch-codecse} lists the MRR scores of the two {\codecse} models on the six programming languages. For each model, the overall score is the average of the six MRR scores. The table shows that $\text{avg\_mlp0\_symmetric}$ has better MRR scores although it has lower alignment scores than $\text{cls\_mlp2\_asymmetric}$. The differences between the two models in MRR scores are small, on average by 0.004. Since the alignment scores consider distances while MRR scores are about ranking, the results indicate that different configurations for {\codecse} models may affect distances among embeddings but do not necessarily change the ranking of the matched pairs.

\subsubsection{Comparing {\codecse} with the baselines}

Table~\ref{tab:eval:codesearch} lists the MRR scores of {\codecse} and all the baselines on the six programming languages. The results show that CodeCSE with zero-shot performs better than CodeBERT and GraphCodeBERT on all the programming languages. \textbf{This indicates that pretraining with contrastive learning is more effective than finetuning for the {\codesearch} task.}

For the models with contrastive learning in pretraining, CodeCSE is slightly better than SynCoBERT and CP3 on average. Specifically, CodeCSE's MRR is higher than SynCoBERT on three programming languages and is higher than C3P on four programming languages. Considering the baselines are finetuned towards specific programming language with (query, code) pairs, \textbf{this result indicates that CodeCSE achieves state-of-the-art results on the {\codesearch} task without fine-tuning, demonstrating the effectiveness of our pre-trained CodeCSE in capturing the semantic similarity between natural language queries and code.}

On the other hand, CodeRetriever performs better than CodeCSE on almost all the cases. The main difference between CodeRetriever's pretraining and CodeCSE's pretraining is that CodeRetriever has a special treatement on unimodal sampling where it uses SimCSE~\cite{gao-2021-simcse} to find positive unimodal samples (Code-Code pairs). The strategy increases the quantity and quality of training samples, which may cause the small difference between CodeRetriever (In-Batch Negatives) and CodeCSE. However, the CodeRetriever model for {\codesearch} is finetuned, so we do not know CodeRetriever's performance without finetuning. As mentioned in Section~\ref{sec:relatedworks:reproducibility}, CodeRetriever is not publically accessible, so no further experiments were conducted on CodeRetriever. 

Furthermore, AR2 boosts CodeRetriever's performance and makes CodeRetriever (AR2) the best approach in this task. This implies that adding advanced finetuning strategies on top of the contrastive learning may improve model performance on the {\codesearch} task. For future extensions, building unimodal datasets and using AR2 are both promising ways for improving {\codecse}.



\subsection{CodeCSE on different programming languages}
The best case for CodeCSE is Ruby, where CodeCSE is the second best peroforming model and its MRR is only 0.001 lower than CodeRetriever (AR2). On the other hand, the bad cases for CodeCSE are Python, Java, and PHP, which are 0.037, 0.038, and 0.036 lower than the best scores. This might be caused by the oversampling strategy we used in pretraining CodeCSE. This strategy makes the model being trained with more iterations on the smaller datasets and fewer iterations on the big datasets. For Ruby (the smallest dataset among the six programming languages), its training dataset size is 24,927, and in each epoch, 49,852 examples are sampled from the training dataset (almost doubled). In contrast, Python is not oversampled (because it is the largest dataset), and Java and PHP are slightly oversampled with ratios of 1.13 and 1.01 respectively.

\section{Threats to Validity}
We used the dataset from {\codesearchnet} out of box and did not examine the quality of the dataset. The differences between the training dataset, validation dataset, and test dataset may affect the generalizability of our conclusions drawn from the case studies on the validation sets.


\section{Conclusion and Future Work}

In this work, we have demonstrated the effectiveness of CodeCSE, which is pretrained on a code-comment dataset (CodeSearchNet~\cite{husain2019codesearchnet}) for getting high-quality code and comment embeddings. However, there is still room for further improvement in future work. Here are some potential directions, and we list them in the order of potential impacts: 
\begin{enumerate}
	\item The use of larger datasets for the self-supervised pretraining stage. While we used CodeSearchNet in this study, which contains millions of (comment, code) pairs, we have observed the success of sentence-BERT~\cite{reimers-2019-sentence-bert} on text sentence embeddings, which was trained on a massive dataset of 1 billion datapoints. Thus, there will be benefits to enlarge the pretraining dataset to improve the quality of the learned code and comment embeddings.
	\item The use of pretrained GPT models to generate embeddings on the enlarged dataset. GPT models have shown great success in natural language processing tasks, and we believe they could be effective in generating high-quality code and comment embeddings. It will be a good choice to explore on the open-sourced versions, such as GPT-Neo~\cite{gpt-neox-library}, GPT-J~\cite{gpt-j}, and BLOOM~\cite{scao2022bloom}, and to report the result comparing with cpt-code, which is not open-sourced.  
	\item SimCSE-style contrastive learning~\cite{gao-2021-simcse} for code and comment embeddings.  SimCSE has shown promising results on text, which doesn't need supervised pairs in contrastive learning. No matter on a unimodal or multimodal dataset, this technique involves learning representations that maximize the similarity between different encoded versions of the same input in a batch. We believe this approach could include more codes that don't have the corresponding comments. 
\end{enumerate}

Overall, these future works have the potential to significantly advance the state-of-the-art in code and comment understanding and can inspire innovative applications in software engineering.

\bibliographystyle{IEEEtran}
\bibliography{main}

\end{document}